\documentclass{webofc}
\usepackage[varg]{txfonts}   

\usepackage[subrefformat=parens]{subcaption}
\usepackage{comment}
\usepackage{xcolor}

\begin{document}
\title
{
\vspace*{-13mm}\hspace{8.41cm} \small{\texttt{YITP-22-105, J-PARC-TH-0276}} \vspace*{7mm} \\
{\fontsize{14pt}{12pt}\selectfont 
Anomalous enhancement of dilepton production 
due to soft modes in dense quark matter
}
}
%
%
\author
{
\firstname{Toru} \lastname{Nishimura}\inst{1, 2}\fnsep\thanks{\email{nishimura@kern.phys.sci.osaka-u.ac.jp}} 
\and
\firstname{Masakiyo} \lastname{Kitazawa}\inst{2, 3, 1}
\and
\firstname{Teiji} \lastname{Kunihiro}\inst{2}
}

\institute
{
Department of Physics, Osaka University, 
560-0043, Toyonaka, Osaka, Japan
\and
Yukawa Institute for Theoretical Physics, Kyoto University, 
606-8502, Kyoto, Japan
\and
J-PARC Branch, KEK Theory Center, Institute of Particle and Nuclear Studies, KEK, 203-1,
319-1106, Shirakata, Tokai, Ibaraki, Japan
}

\abstract
{
We explore how the dilepton production rate is modified 
near the critical temperature of color superconductivity and QCD critical point
by the soft modes inherently associated with the phase transitions of second-order.
It is shown that the soft modes affect the photon self-energy 
significantly through so called the Aslamasov-Larkin, Maki-Thompson and density of states terms,
which are known responsible for the paraconductivity 
in the metalic superconductivity, and cause an anomalous enhancement of 
the production rate in the low energy/momentum region.
}

\maketitle

\section{Introduction}
\label{Introduction}
Rich physics is expected to exist in high baryon-density matter at finite temperature,
and may be hopefully revealed by experimental programs in relativistic heavy-ion collisions (HIC) 
such as the beam energy scan program at RHIC, HADES and NA61/SHINE, 
as well as the future plans at FAIR, NICA and J-PARC-HI. 
In this report, we show a promising observability of 
the color superconducting (CSC)~\cite{Alford:2007xm} phase transition
and QCD critical point (CP) in these experiments 
through an anomalous enhancement of the dilepton production rate (DPR) 
caused by the soft modes inherently associated with these
phase transitions of second-order;
the CSC phase transion is assumed to be second-order~\cite{Nishimura:2022mku}.
We calculate the effects of the soft modes 
on the DPR by extending the theory of the paraconductivity in metals~\cite{book_Larkin},
and show that the soft modes lead to an anomalous enhancement of the DPR 
at low energy region near each phase transition.

\section{Formalism}
\label{Formalism}
To explore the temperature and density region around the 2-flavor color superconductivity (2SC) 
and QCD CP, which is expected to be realized at relatively low densities, 
we employ the 2-flavor NJL model
\begin{align}
  \mathcal{L} &= \bar{\psi} i ( \gamma^\mu \partial_\mu - m ) \psi 
  + \ G_{\rm S} [(\bar{\psi} \psi)^2 + (\bar{\psi} i \gamma_5 \vec{\tau} \psi)^2]
  + \ G_{\rm C} (\bar{\psi} i \gamma_5 \tau_2 \lambda_A \psi^C)(\bar{\psi}^C i \gamma_5 \tau_2 \lambda_A \psi),
  \label{eq:lagrangian}
\end{align}
where $\tau_2$ ($\gamma_{A=2,5,7}$) is the antisymmetric component 
of the Pauli (Gell-Mann) matrices for the flavor $SU(2)_f$ (color $SU(3)_c$), and 
$\psi^C (x) \equiv C \bar{\psi}^T (x)$ with $C = i \gamma_2 \gamma_0$.
We consider the chiral limit for the analysis of the DPR near the critical temperature ($T_c$) 
of the 2SC,
while we set to $m=4$~MeV for investigating the QCD CP.
The scalar coupling constant $G_S=5.01~\rm{MeV^{-2}}$ and the
three-momentum cutoff $\Lambda=650$~MeV are
used, which leads to the pion decay constant 
$f_{\pi}=93~\rm{MeV}$ and the chiral condensate 
$\langle \bar{\psi} \psi \rangle = (-250~\rm{MeV})^3$,
while the diquark coupling $G_C$ is treated as a free parameter.

The DPR is related to the retarded photon self-energy $\Pi^{R \mu\nu}(k)$ as 
\begin{align}
\frac{d^4\Gamma}{d^4k} = - \frac{\alpha}{12\pi^4} 
\frac{1}{k^2} \frac{1}{e^{\omega/T}-1} g_{\mu\nu}  
{\rm Im} \Pi^{R \mu\nu} (k) ,\label{eq:RATE-kom}
\end{align}
where $k=(\mathbf{k}, \omega)$ is the four momentum
of the photon and $\alpha$ is the fine structure constant. 
In this study, we calculate the modification of $\tilde{\Pi}^{\mu\nu}(k)$
due to the soft modes of the 2SC and QCD CP.
In this section, we first calculate $\tilde{\Pi}^{\mu\nu}(k)$ including 
the diquark soft modes of 2SC~\cite{Nishimura:2022mku} in Sec.~\ref{diquark},
and then deal with the soft modes of the QCD CP that are the mesonic modes
in Sec.~\ref{mesonic}.

\subsection{Modification of $\Pi^{\mu\nu}$ by diquark soft modes}
\label{diquark}
It is known that the diquark fluctuations form 
a collective mode with a significant strength 
above but near $T_c$ of the 2SC~\cite{Kitazawa:2005vr, Kunihiro:2007bx}.
The collective mode associated with the second-order nature of
phase transition is called the soft mode.
The propagator of  the diquark soft modes in the random-phase approximation (RPA)
in the imaginary-time formalism is given by
\begin{align}
  \tilde{\Xi}_C(k) = \frac{1}{{G_C}^{-1}+\mathcal{Q}_C(k)},~~~~
  \mathcal{Q}_C (k) = 
   \begin{minipage}[h]{0.12\linewidth}
    \centering
    \includegraphics[keepaspectratio, scale=0.05]{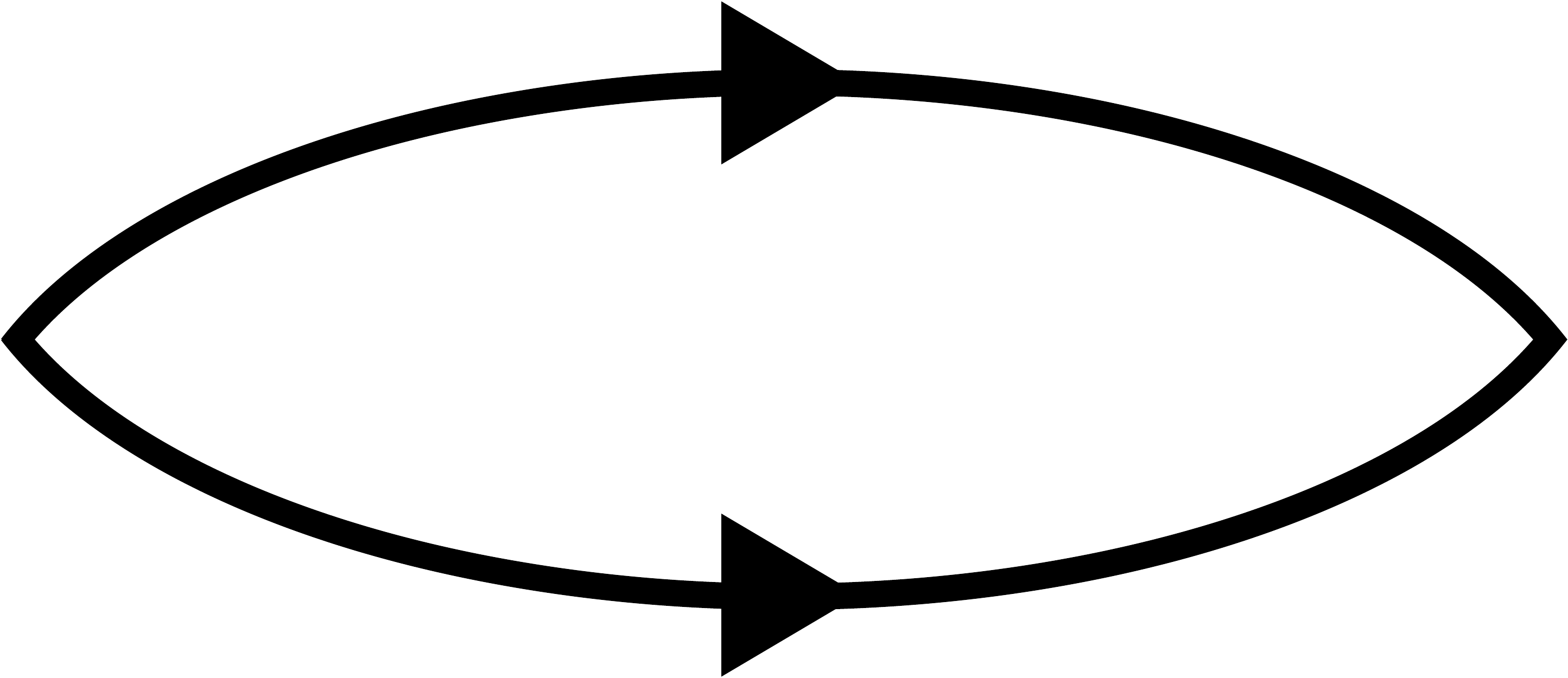}
   \end{minipage}
  =-8 \int_p {\rm Tr} [\mathcal{G}_0 (k-p) \mathcal{G}_0(p)],
  \label{eq:softmode_CSC} 
\end{align}
where $\mathcal{Q}_C(k)=\mathcal{Q}_C(\mathbf{k}, i\nu_n)$ is 
the one-loop $qq$ correlation function with
$\mathcal{G}_0(p)=\mathcal{G}_0(\mathbf{p}, i\omega_m) = 
1/[(i\omega_m + \mu)\gamma_0 - \mathbf{p} \cdot \mathbf{\gamma}]$ 
the free quark propagator, 
$\omega_m$ ($\nu_n$)  the Matsubara frequency for fermions (bosons), and
${\rm Tr}$ the trace over the Dirac indices.
The retarded propagator is obtained 
by the analytic continuation 
$\Xi^R_C (\mathbf{k}, \omega) = \tilde{\Xi}_C (\mathbf{k}, i\nu_n \rightarrow \omega+i\eta)$.

We remark that $\Xi^R_C (\mathbf{k}, \omega)$ satisfies the Thouless criterion, that is
$[\Xi^R_C (\mathbf{0}, 0)]^{-1} = 0$ at $T=T_c$ for the second-order phase transition.
The criterion shows that $\Xi^R_C (\mathbf{k}, \omega)$
has a pole at the origin at $T=T_c$,
and hence the diquark fluctuations become soft near $T_c$~\cite{Kitazawa:2005vr}.
This fact also allows us to adopt the time-dependent Ginzburg-Landau (TDGL) approximation 
\begin{align}
[\Xi^R_C(\mathbf{k}, \omega)]^{-1} = A(\mathbf{k}) + C(\mathbf{k}) \omega, \label{eq:softmodeapprox}
\end{align}
near but above $T_c$, where $A(\mathbf{k}) = G_C^{-1} + Q^R_C (\mathbf{k}, 0)$ 
and $C(\mathbf{k}) = \partial Q^R_C(\mathbf{k}, \omega)/\partial \omega~|_{\omega=0}$.
It is known that Eq.~(\ref{eq:softmodeapprox}) reproduces $\Xi^R_C(\mathbf{k}, \omega)$
over wide ranges of $\omega$, $\mathbf{k}^2$ and $T(>T_c)$~\cite{Kitazawa:2005vr}.

\begin{figure}[t]
    \begin{tabular}{cccc}
      \begin{minipage}[t]{0.212\hsize}
        \centering
        \includegraphics[keepaspectratio, scale=0.09]{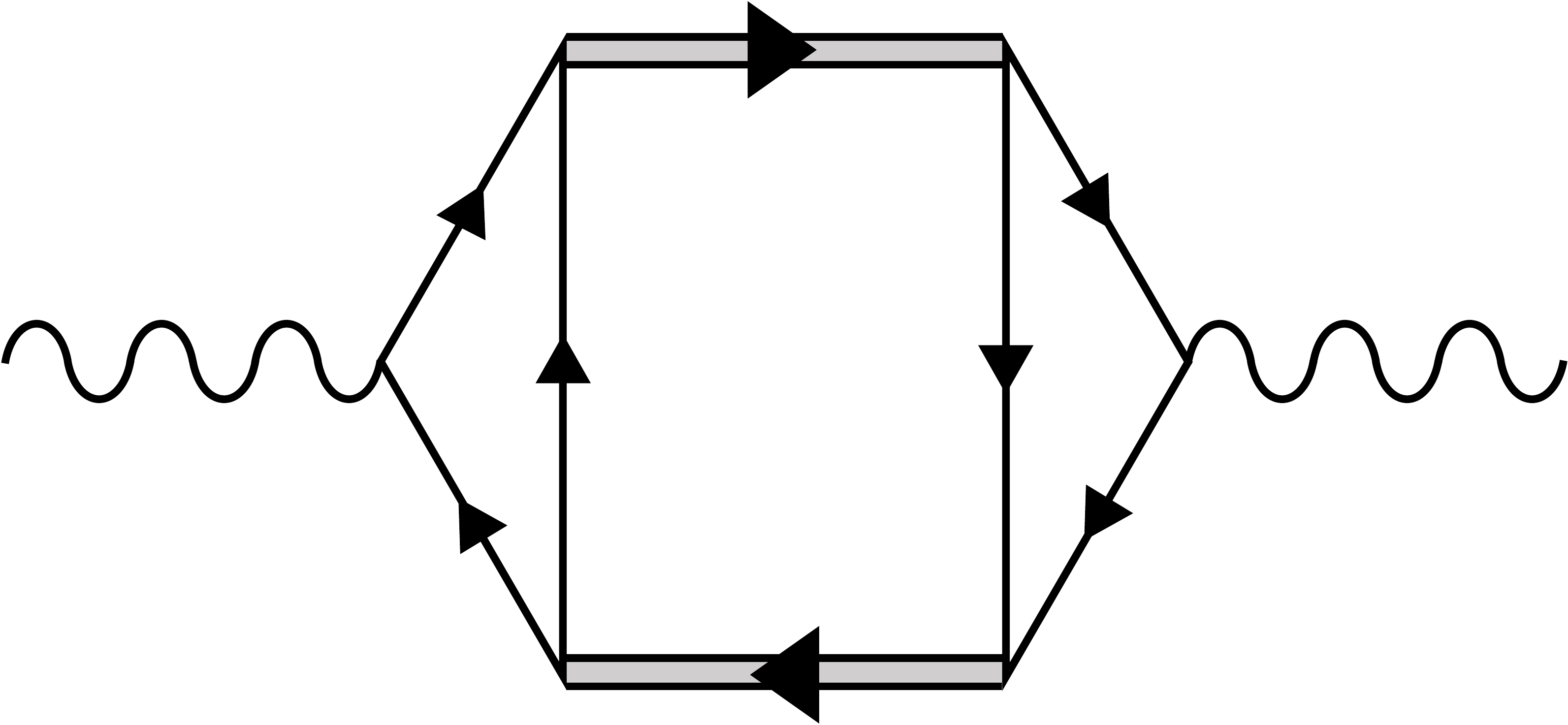}
        \subcaption{}
        \label{fig_AL}
      \end{minipage} &
      \begin{minipage}[t]{0.212\hsize}
        \centering
        \includegraphics[keepaspectratio, scale=0.09]{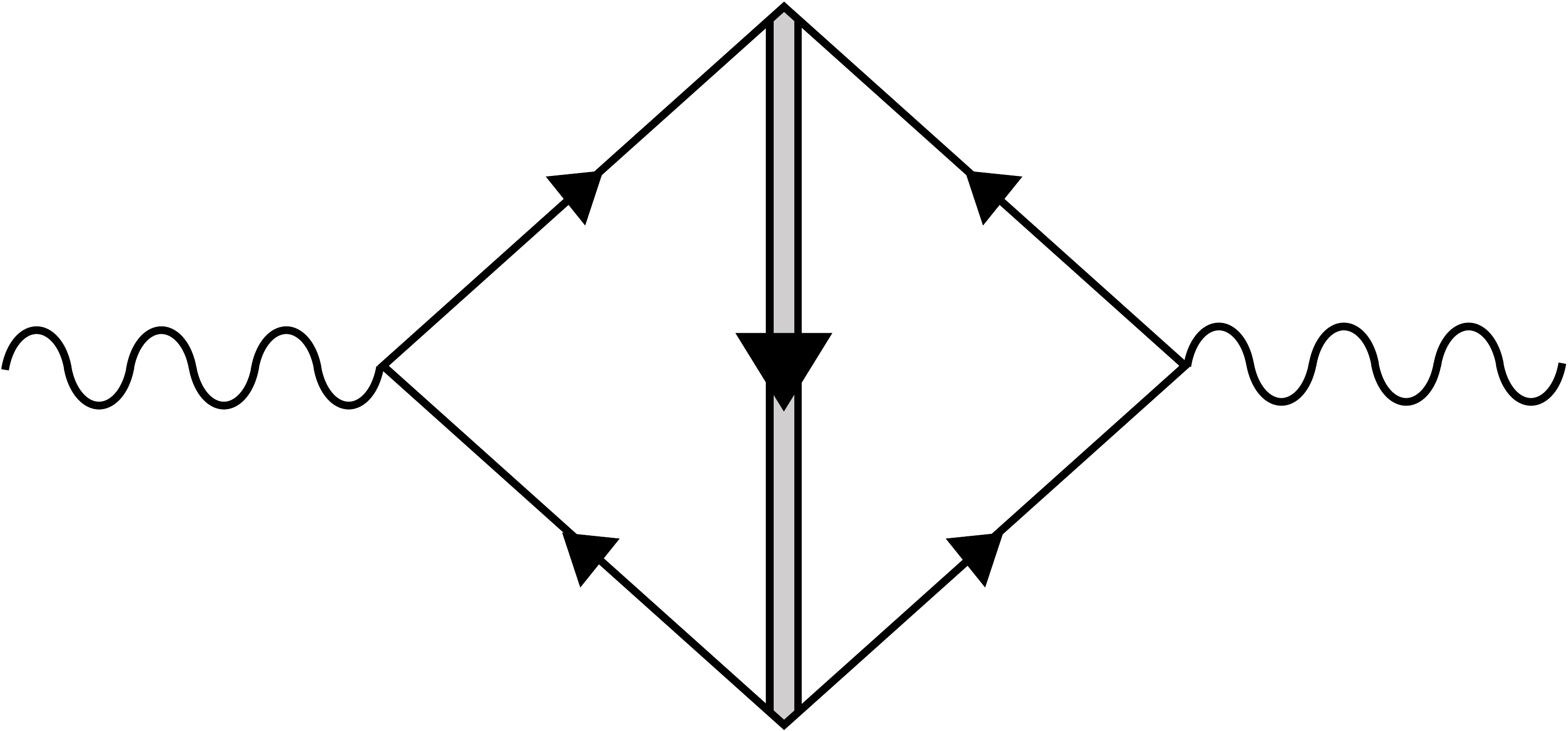}
        \subcaption{}
        \label{fig_MT}
      \end{minipage} &
      \begin{minipage}[t]{0.212\hsize}
        \centering
        \includegraphics[keepaspectratio, scale=0.09]{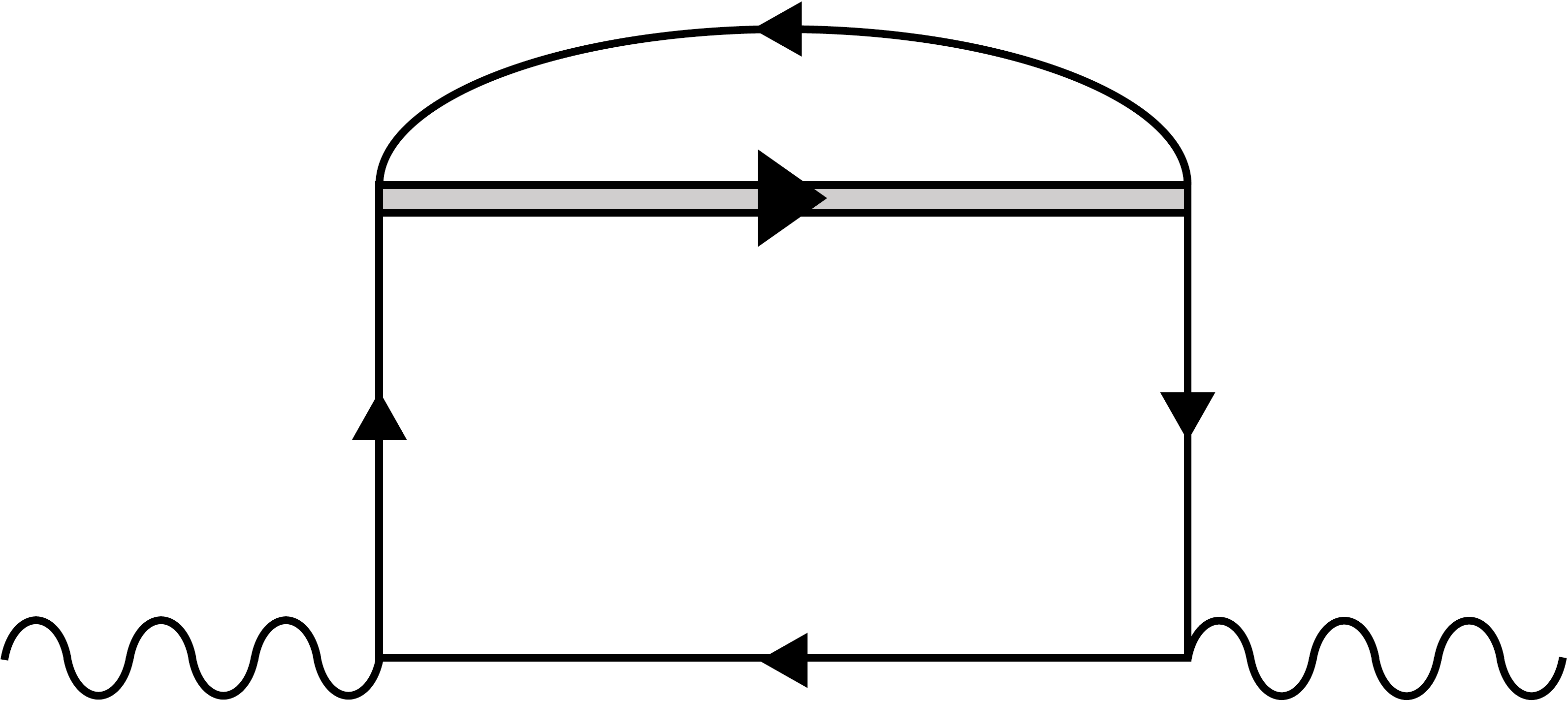}
        \subcaption{}
        \label{fig_DOS1}
      \end{minipage} &
      \begin{minipage}[t]{0.212\hsize}
        \centering
        \includegraphics[keepaspectratio, scale=0.09]{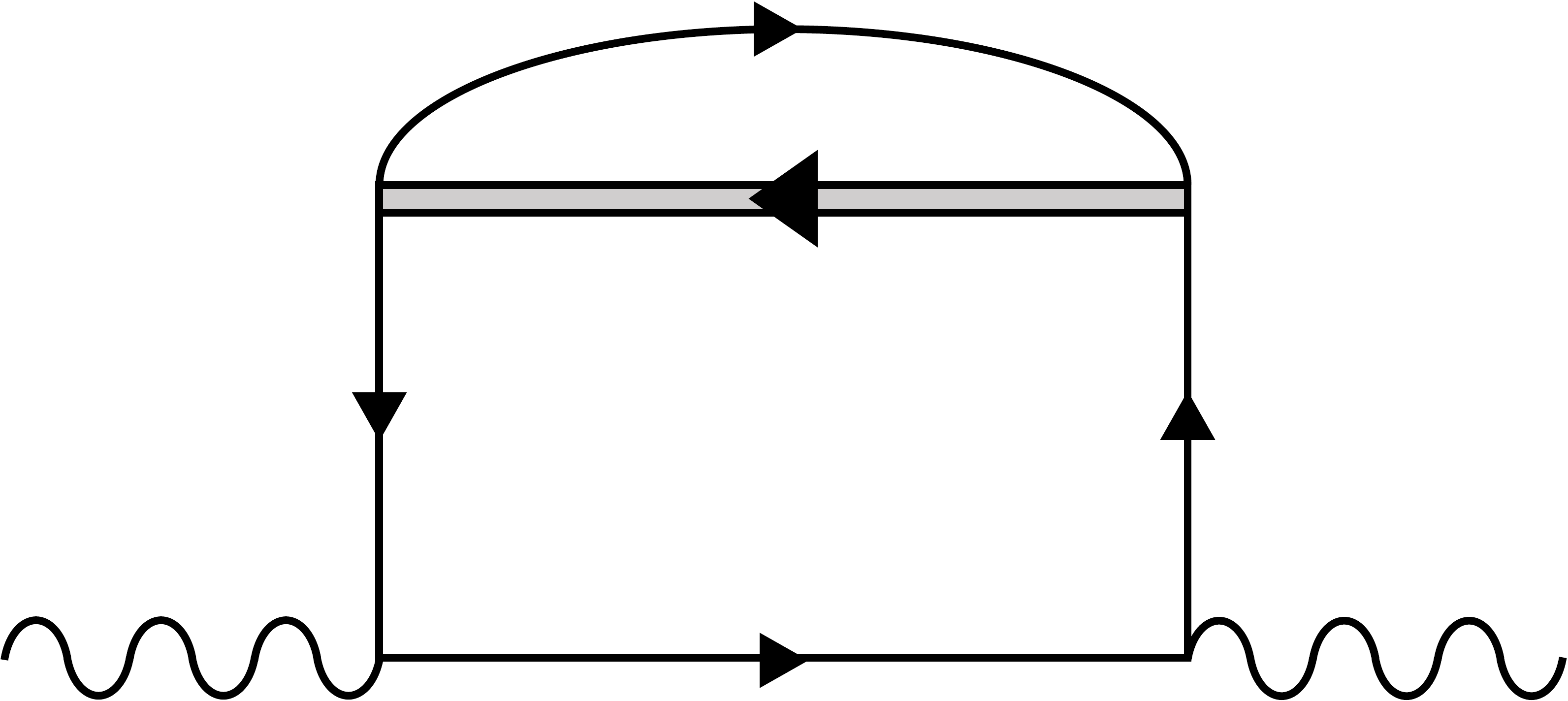}
        \subcaption{}
        \label{fig_DOS2}
      \end{minipage} 
    \end{tabular}
    \vspace*{-0.65cm}
    \caption{Diagrammatic representations of the Aslamasov-Larkin (a),
      Maki-Thompson (b) and density of states (c,d) terms with the diquark soft mode.
      The wavy and double lines are photons and the diquark soft modes, respectively.}
    \vspace*{-0.7cm}
     \label{fig:selfenergy}
\end{figure}

To construct $\tilde{\Pi}^{\mu \nu} (k)$ that inolves the soft mode,
we start from the one-loop diagram of $\tilde{\Xi}_C(\mathbf{k}, i\nu_n)$, 
which is the lowest contribution of
the soft modes to the thermodynamic potential~\cite{Nishimura:2022mku}.
The photon self-energy is then constructed by attaching 
electromagnetic vertices at two points of quark lines in the thermodynamic potential.
One then obtains four types of diagrams 
shown in Fig.~\ref{fig:selfenergy}.
These diagrams are called the Aslamasov-Larkin (AL) 
(Fig.~\ref{fig:selfenergy}(a)), 
Maki-Thompson (MT) (Fig.~\ref{fig:selfenergy}(b))
and density of states (DOS) (Fig.~\ref{fig:selfenergy}(c, d))~\cite{book_Larkin} 
terms, respectively,
in the theory of metallic superconductivity.
Each contribution to the photon self-energy, 
$\tilde{\Pi}_{\rm AL}^{\mu\nu} (k)$, $\tilde{\Pi}_{\rm MT}^{\mu\nu} (k)$ and $\tilde{\Pi}_{\rm DOS}^{\mu\nu} (k)$, 
respectively, is given by
\begin{align}
\tilde{\Pi}_{\rm AL}^{\mu\nu} (k) &= 3 \int_q 
\tilde{\Gamma}^\mu(q, q+k) \tilde{\Xi}_C(q+k) \tilde{\Gamma}^\nu(q+k, q) \tilde{\Xi}_C(q),
\label{eq:AL} \\
\tilde{\Pi}_{\rm MT (DOS)}^{\mu\nu} (k) &= 3 \int_q 
\tilde{\Xi}_C(q) \mathcal{R}_{\rm MT (DOS)}^{\mu\nu}(q, k), \label{eq:MTDOS} 
\end{align}
where $\tilde{\Gamma}^\mu(q, q+k)$ and 
$\mathcal{R}^{\mu\nu}(q, k)=\mathcal{R}_{\rm MT}^{\mu\nu}(q, k)+\mathcal{R}_{\rm DOS}^{\mu\nu}(q, k)$ 
satisfy the following Ward-Takahashi (WT) identities 
\begin{align}
k_\mu \tilde{\Gamma}^\mu (q, q+k) &= (e_u+e_d) [\mathcal{Q}_C(q+k)-\mathcal{Q}_C(q)], 
\label{eq:AL-vertex_CSC} \\
k_\mu \mathcal{R}^{\mu\nu} (q, k) &= (e_u+e_d) [\tilde{\Gamma}^\nu (q-k, q)-\tilde{\Gamma}^\nu (q, q+k)], 
\label{eq:MTDOS-vertex_CSC}
\end{align}
where $e_u=2|e|/3$ ($e_d=-|e|/3$) is the electric charge of up (down) quarks.
The total photon self-energy is then given by
$\tilde{\Pi}^{\mu\nu} (k) = \tilde{\Pi}_{\rm free}^{\mu\nu} (k) 
+\tilde{\Pi}_{\rm AL}^{\mu\nu} (k)+\tilde{\Pi}_{\rm MT}^{\mu\nu} (k)+\tilde{\Pi}_{\rm DOS}^{\mu\nu} (k) $. 
One can explicitly check that this photon self-energy satisfies 
the WT identity using Eqs.~(\ref{eq:AL-vertex_CSC}) and (\ref{eq:MTDOS-vertex_CSC}).

To obtain the DPR with Eq.~(\ref{eq:RATE-kom}),
the calculation of the spatial components of $\Pi^{R\mu\nu}(k)$ is
sufficient since $\Pi^{R00}(k)$ is expressed in terms of the longitudinal
part from the WT identity as $\Pi^{R00} (k) = \mathbf{k}^2\Pi^{R11} (k)/k_0^2$
with $k=(k_0, |\mathbf{k}|, 0, 0)$.
For the vertex functions (\ref{eq:AL-vertex_CSC}) and (\ref{eq:MTDOS-vertex_CSC}), 
we approximate $\tilde{\Gamma}^\mu(q, q+k)$ and $\mathcal{R}^{\mu\nu}(q, k)$ using
Eq.~(\ref{eq:softmodeapprox}).
For $\tilde{\Gamma}^i(q, q+k)$, we employ an ansatz
\begin{align}
\tilde{\Gamma}^i (q, q+k) =  -(e_u+e_d) 
\frac{\mathcal{Q}_C(\mathbf{q}+\mathbf{k}, 0)-\mathcal{Q}_C(\mathbf{q}, 0)}
{|\mathbf{q}+\mathbf{k}|^2 - |\mathbf{q}|^2} (2q+k)^i, 
\label{eq:AL-vertex-approx}
\end{align}
which is a real number and satisifes the WT identity~(\ref{eq:AL-vertex_CSC}) with Eq.~(\ref{eq:softmodeapprox}).
One can also obtain $\mathcal{R}^{ij}(q, k)$ in the same manner, and finds that it is a real number as well.
Using this vertex and Eq.~(\ref{eq:softmodeapprox}),
the imaginary part of $\Pi^{Rij}_{\rm MT}(q)+ \Pi^{Rij}_{\rm DOS}(q)$ vanishes. 
Therefore, the MT and DOS terms do not contribute to the calculation of the DPR.
This result is in accordance with
the case of the metallic superconductivity~\cite{book_Larkin}.

\subsection{Modification of $\Pi^{\mu\nu}$ by mesonic soft modes near QCD CP}
\label{mesonic}
The photon self-energies including the mesonic soft modes of the QCD CP can also be calculated with a similar manner.
The propagator of the mesonic mode is  evaluated by the RPA as
\begin{align}
  \tilde{\Xi}_S(k) = \frac{1}{{G_S}^{-1}+\mathcal{Q}_S(k)},~~~~
  \mathcal{Q}_S (k) = 
   \begin{minipage}[h]{0.12\linewidth}
    \centering
    \includegraphics[keepaspectratio, scale=0.05]{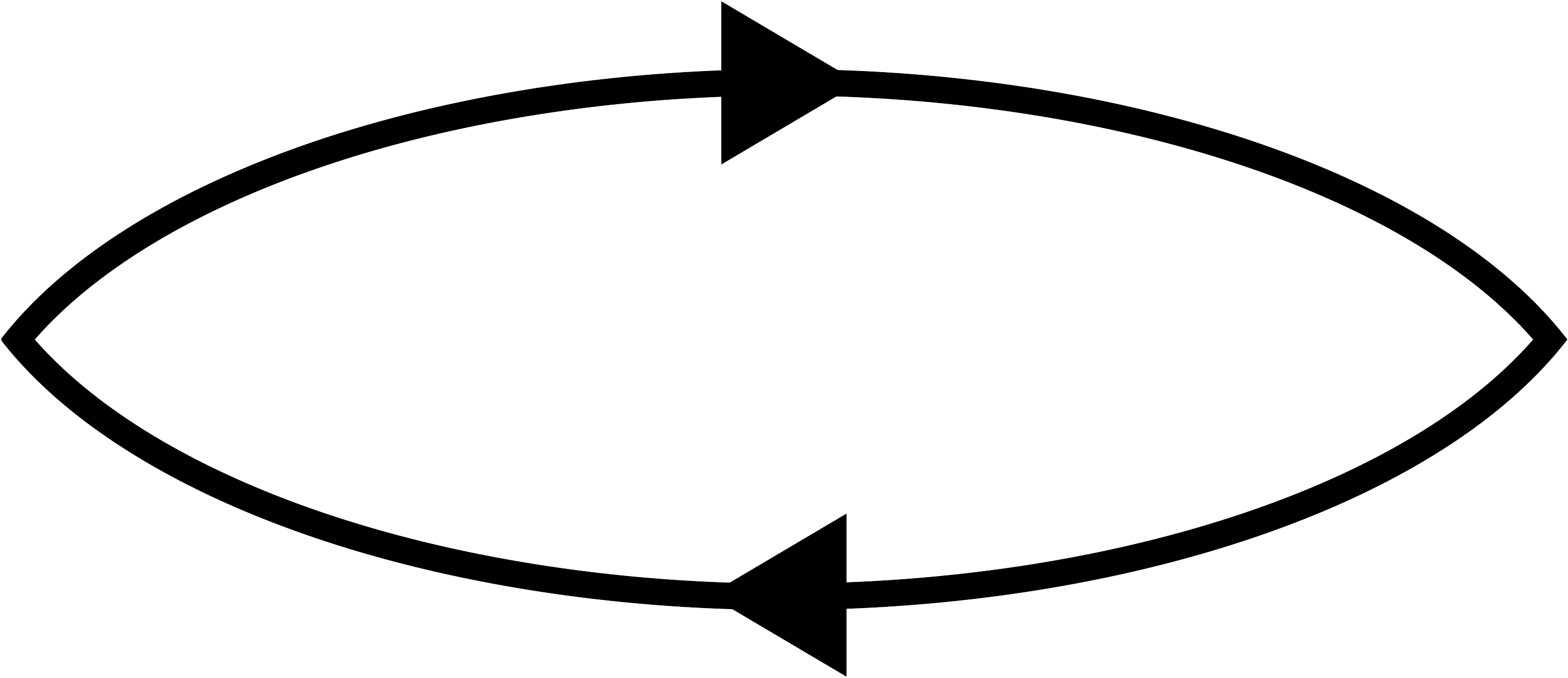}
    \end{minipage}
  =12 \int_p {\rm Tr} [\mathcal{G}_0 (p-k) \mathcal{G}_0(p)],
  \label{eq:softmode_QCDCP} 
\end{align}
where $\mathcal{Q}_S (k)$ is the one-loop $q \bar{q}$ correlation function. 
As described in Ref.~\cite{Fujii:2004jt}, $\Xi^R_S(k)$ has a discontinuity 
at the light cone, 
and its strength is limited to the space-like region.
This property is contrasted with the diquark soft mode~(\ref{eq:softmode_CSC}), which 
does not have the discontinuity at the light cone, while the diquark soft mode also 
has a significant strength in the space-like region.
The photon self-energy with the mesonic modes is given by
\begin{align}
\tilde{\Pi}_{\rm AL}^{\mu\nu} (k) &= \sum_f \int_q 
\tilde{\Gamma}^\mu_f(q, q+k) \tilde{\Xi}_S(q+k) \tilde{\Gamma}^\nu_f(q+k, q) \tilde{\Xi}_S(q), 
\label{eq:AL_QCDCP} \\
\tilde{\Pi}_{\rm MT (DOS)}^{\mu\nu} (k) &= \sum_f \int_q 
\tilde{\Xi}_S(q) \mathcal{R}_{f~\rm MT (DOS)}^{\mu\nu} (q, k), 
\label{eq:MTDOS_QCDCP} 
\end{align}
where $f=u, d$ are the indices of the flavors.
$\tilde{\Gamma}^\mu_f (q, q+k)$ and 
$\mathcal{R}^{\mu\nu}_f (q, k)=\mathcal{R}_{f~{\rm MT}}^{\mu\nu}(q, k)+\mathcal{R}_{f~{\rm DOS}}^{\mu\nu}(q, k)$ 
satisfy the WT identity,
similarly to Eqs.~(\ref{eq:AL-vertex_CSC}) and (\ref{eq:MTDOS-vertex_CSC}).

\section{Numerical results and summary}
\label{Numerical results and summary}

\begin{figure}[tbp]
\vspace*{-0.4cm}
	\begin{tabular}{cc}
      \begin{minipage}[t]{0.48\hsize}
        \centering
        \includegraphics[keepaspectratio, scale=0.4]{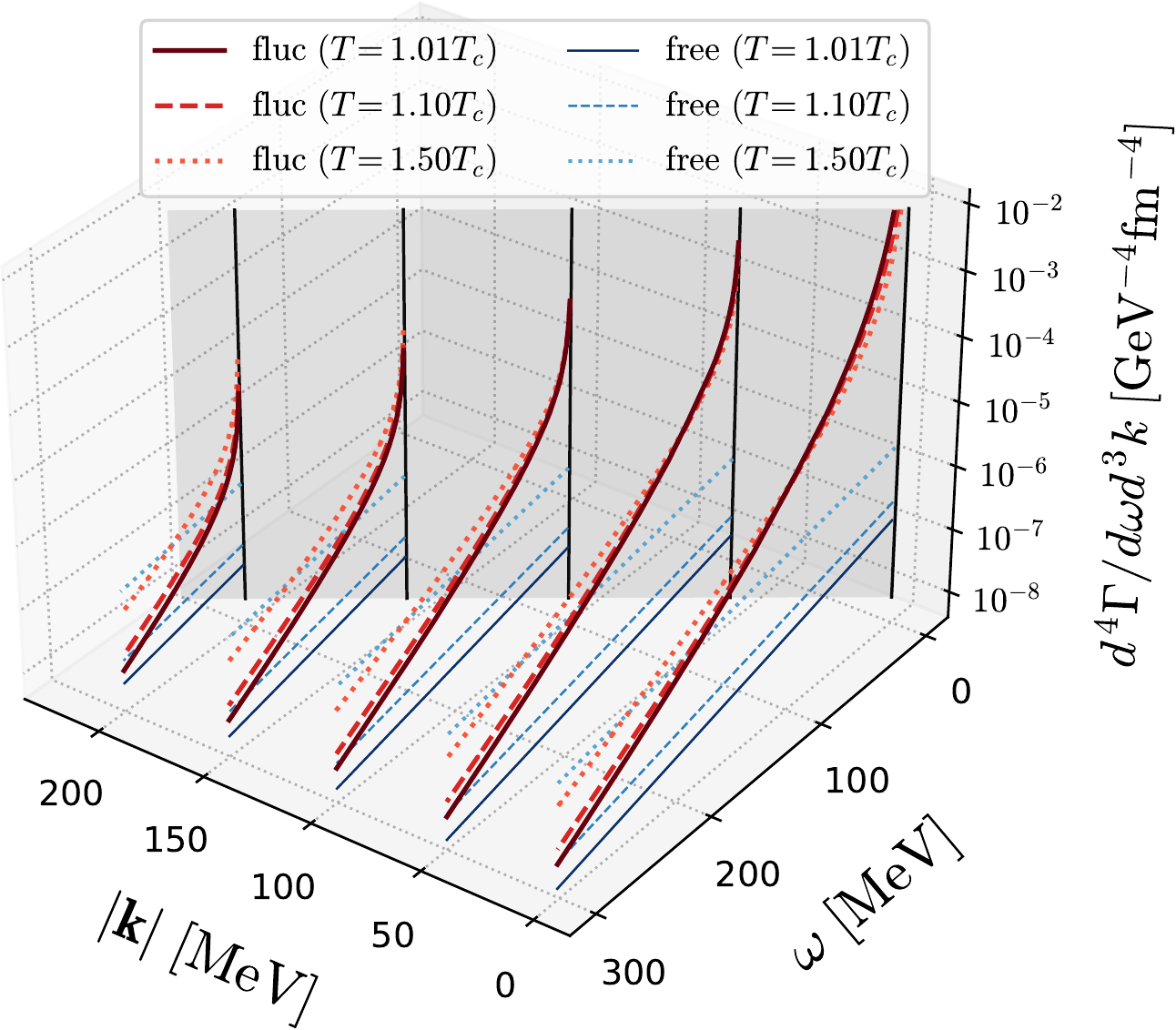}
      \end{minipage} &
      \begin{minipage}[t]{0.48\hsize}
        \centering
        \includegraphics[keepaspectratio, scale=0.4]{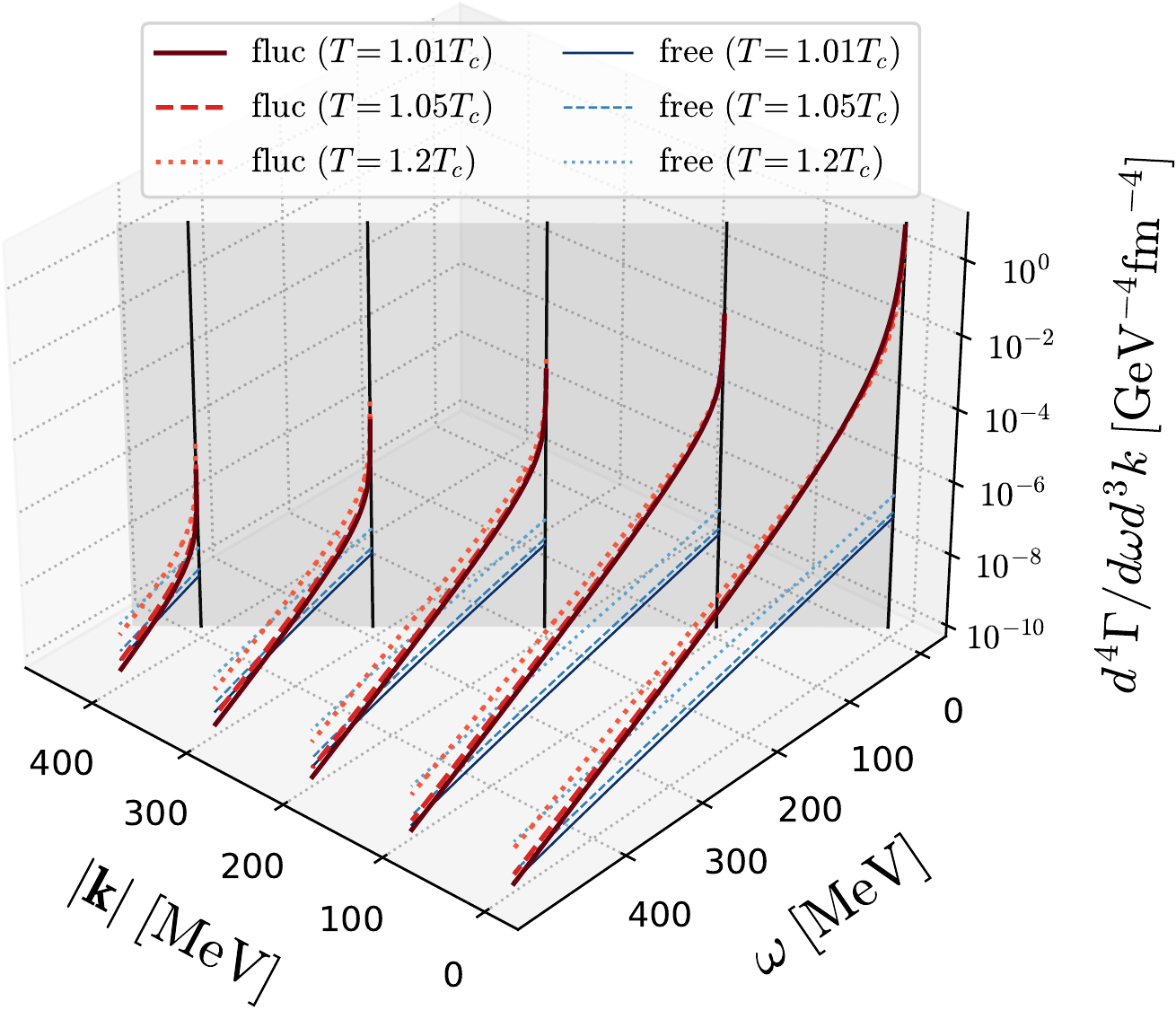}
      \end{minipage}
	\end{tabular}
 \vspace*{-0.3cm}
\caption{
DPRs per unit energy $\omega$ and momentum $\mathbf{k}$
above $T_c$ of the 2SC at $\mu = 350~{\rm MeV}$ (left)~\cite{Nishimura:2022mku}
and of the QCD CP at $\mu = \mu_{\rm CP}$ (right). 
The thick lines are the contribution of the soft modes, 
and the thin lines are the results for the free quark gas. 
}
\vspace*{-0.7cm}
\label{fig:RATE}
\end{figure}

In the left panel of Fig.~\ref{fig:RATE}, we show the DPR
per unit energy and momentum $d^4 \Gamma / d^4k$
near the 2SC phase transition
for $T=1.01 T_c$, $1.1 T_c$ and $1.5 T_c$ at $\mu=350~{\rm MeV}$ 
($T_c =42.94~{\rm MeV}$)~\cite{Nishimura:2022mku}.
The thick lines are the contribution from the soft modes, 
while the thin lines are the ones of the free quark gas.
One sees that the DPR from the
soft modes is anomalously enhanced 
in the small $\omega$ and $\mathbf{k}$ region in comparison 
with the free quark gas for $T\lesssim1.5T_c$,
and this enhancement becomes more pronounced as $T$ approaches $T_c$.
This result is expected through the properties of the soft modes.
Shown in the right panel of Fig.~\ref{fig:RATE} is the DPR near the QCD CP
for $T=1.01 T_{\rm CP}$, $1.05 T_{\rm CP}$ and $1.2 T_{\rm CP}$ at $\mu=\mu_{\rm CP}$,
where $(T_{\rm CP}, \mu_{\rm CP}) = (35.07,322.20)~{\rm MeV}$ is the location of the QCD CP.
One finds that the anomalous enhancement of the DPR is also observed 
similarly to the case of the 2SC.

The low-energy part of the DPR is related to the electrical conductivity $\sigma$ and relaxation time $\tau$,
which are obtained from differentiations of 
$\rho_L(\mathbf{0}, \omega)=
\sum_i {\rm Im} {\Pi_{AL}^R}^{ii} (\mathbf{0}, \omega)/\pi$ at $\omega=0$.
The analysis of $\sigma$ and $\tau$ thus allows us to understand behavior of the DPR near $T_c$ more quantitatively.
We find that these quantities diverge at $T_c$ with
$\sigma \propto \epsilon^{-1/2}$ and $\tau \propto \epsilon^{-3/2}$ for the 2SC,
while for the case of the QCD CP, $\sigma \propto \epsilon^{-2/3}$ and $\tau \propto \epsilon^{-1}$ 
with $\epsilon=|T-T_c|/T_c$, as will be discussed in the forthcoming publication.
The origin of the difference of two cases is due to the $T$-dependence of $A(\mathbf{k})$ 
and the $\mathbf{k}$-dependence of  $C(\mathbf{k})$ in Eq.~(\ref{eq:softmodeapprox}).

In this report, we studied how the soft modes of the 2SC and QCD CP
affect the DPR near but above $T_c$.
The modification of the photon self-energy due to the soft modes
are investigated through the analysis of 
the Aslamasov-Larkin, Maki-Thompson and density of states terms
with the TDGL approximation for the soft mode propagator and vertices.
The enhancement of the DPR found in this study
would allow us to detect these signals of the 2SC and QCD CP in the future HIC experiments.






%
%
%


\end{document}